\documentclass[preprint,12pt]{elsarticle}
\usepackage{amssymb}
\usepackage{mathrsfs}
\usepackage{slashbox}     
\usepackage{epstopdf}
\usepackage{amsthm}
\usepackage{caption2}                   
\renewcommand{\figurename}{Fig.}        

\usepackage{xcolor}   
\usepackage[colorlinks, citecolor=blue]{hyperref}
\definecolor{emph_color}{rgb}{0,0,1}  
\hypersetup{colorlinks=false}                       
\theoremstyle{plain}

\theoremstyle{definition}

\journal{arxiv.org}

\begin{document}
\title{On optimal tempered L{\'e}vy flight foraging}


\author[ustc]{Yuquan Chen}
\author[unsw]{Derek Hollenbeck}
\author[ustc]{Yong Wang}
\author[unsw]{YangQuan Chen\corref{cor1}}
\ead{yongwang@ustc.edu.cn}
\address[ustc]{Department of Automation, University of Science and Technology of China, Hefei 230026, China}
\address[unsw]{Mechatronics, Embedded Systems and Automation Lab, University of California, Merced, CA 95343, USA}

\cortext[cor1]{Corresponding author}

%
%
%

\begin{abstract}
Optimal random foraging strategy has gained increasing concentrations. It is shown that L{\'e}vy flight is more efficient compared with the Brownian motion when the targets are sparse. However, standard L{\'e}vy flight generally cannot be followed in practice. In this paper, we assume that each flight of the forager is possibly interrupted by some uncertain factors, such as obstacles on the flight direction, natural enemies in the vision distance, and restrictions in the energy storage for each flight, and introduce the tempered L{\'e}vy distribution $p(l)\sim {\rm e}^{-\rho l}l^{-\mu}$. It is validated by both theoretical analyses and simulation results that a higher searching efficiency can be derived when a smaller $\rho$ or $\mu$ is chosen. Moreover, by taking the flight time as the waiting time, the master equation of the random searching procedure can be obtained. Interestingly, we build two different types of master equations: one is the standard diffusion equation and the other one is the tempered fractional diffusion equation.
\end{abstract}

\maketitle



\section{Introduction}
One common approach to the animal movement patterns is to use the scheme of optimizing random search \citep{bartumeus2005animal,viswanathan2008Levy,reynolds2008optimal}. In a random search model, single or multiple individuals search a landscape to find targets whose locations are not known \emph{a priori}, which is usually adopted to describe the scenario of animals foraging for food, prey or resources. The locomotion of the individual has a certain degree of freedom which is characterized by a specific search strategy such as a type of random walk and is also subject to other external or internal constraints, such as the environmental context of the landscape or the physical and psychological conditions of the individual. It is assumed that a strategy that optimizes the search efficiency can evolve in response to
such constraints on a random search, and the movement is a consequence of the optimization on random search.

Many researchers have concentrated on the study of different animals' foraging movements. It is shown that when the environment contains a high density of food items, foragers tend to adopt Brownian walks, characterized by a great number of short step lengths in random directions that maintain foragers in a small portion of the available space \citep{bartumeus2002optimizing,kerster2016spatial}. In contrast, when the density of food items is low, individuals tend to exhibit L{\'e}vy flights, where larger step lengths occasionally occur and relocate the foragers in the environment. Due to the fact that the density of food items is often low, many animals behave a L{\'e}vy flight when foraging and their movements have been found to fit closely to a L{\'e}vy distribution (power law distribution) with an exponent close to 2 \citep{viswanathan1999optimizing,viswanathan2000Levy}. For instance, the foraging behavior of the wandering albatross on the ocean surface was found to obey a power law distribution \citep{viswanathan1996Levy}; the foraging patterns of a free-ranging spider monkey in the forests was also found to be a power law tailed distribution of steps consistent with L{\'e}vy walks \citep{boyer2004modeling,ramos2004Levy}.

On this basis, researchers mainly consider two issues: one is to model the foraging behavior as a L{\'e}vy flight and the other one is to study the searching efficiency theoretically or experimentally. It is assumed that the forager behaves a random walk according to the distribution $p(l)\sim l^{-\mu},~1<\mu<3$. Then it is proven that the highest searching efficiency can be obtained when $\mu$ is close to 2 for the non-destructive case. While the searching efficiency is higher when $\mu$ tends to 1 for the destructive case. Later, many more complex situations are considered. Due to the fact that foragers are always searching in a bounded area, \cite{dybiec2017Levy} and \cite{zhao2015optimal} studied the searching efficiency of L{\'e}vy flight in a bounded area. \cite{kerster2016spatial} took the spatial memory of foragers into consideration and concluded that the spatial information influenced the foraging behavior significantly according to the experimental results. Interestingly, it was claimed that the L{\'e}vy flight foraging behavior can also be interpreted by a composite search model \citep{plank2008optimal,bartumeus2014stochastic}. The model consists of an intensive search phase, followed by an extensive phase, if no food is found in the intensive phase. Particularly, \cite{zeng2014optimal} considered the waiting time between two successive flights and formulated the master equation for such foraging behavior.

Though many studies have proven that it is usually more efficient to utilize L{\'e}vy flight foraging strategy, standard L{\'e}vy flight cannot be followed in practice because of many uncertain factors. For instance, the forager may encounter obstacles or natural enemies and extremely large flight distance cannot be reasonable due to the restriction of the forager's flight ability. In this paper, we take these conditions into consideration and temper the L{\'e}vy distribution with an exponential decaying function, which results in a tempered L{\'e}vy distribution $p(l)\sim {\rm e}^{-\rho l}l^{-\mu}$. It is then shown that a higher searching efficiency will be derived when a smaller $\rho$ or $\mu$ is chosen, both by simulation and theoretical analyses. Further, two different types of master equations are derived: one is the standard diffusion equation and the other one is the tempered fractional diffusion equation. Since the first and second order moments exist, the foraging movement will finally result in a Gaussian motion, which indicates that the tempered fractional diffusion equation is in fact another expression for the standard diffusion.

The remainder of the paper is organized as follows. Section \ref{sec2} provides the basic foraging model and some basic results are also given. In Section \ref{sec3}, we study the searching efficiency when a tempered L{\'e}vy distribution is considered. Two different types of  master equations are derived in Section \ref{sec4} after treating the flight time as the waiting time. The paper is concluded in Section \ref{sec5}.

\section{ Basic definitions and model description}\label{sec2}
In this section, we mainly recall the original model and basic results of L{\'e}vy flight optimal random search. Assume that target sites are uniformly distributed and the forager behaves as follows
\begin{itemize}
  \item [(1)] If a target site lies within a ``direct vision" distance $r_v$, then the forager moves on a straight line to the nearest site. A finite value of $r_v$, no matter how large, models the constraint that no forager can detect a target site located an arbitrarily large distance away.
  \item [(2)] If there is no target site within a distance $r_v$, then the forager chooses a direction randomly and a distance $l_j$ from a probability distribution. It then incrementally moves to the new point, constantly looking for a target within a radius $r_v$ along its way. If it does not detect a target, it stops after traversing the distance $l_j$ and chooses a new direction and a new distance $l_{j+1}$; otherwise, it proceeds to the target as rule (1).
\end{itemize}
In the case of non-destructive foraging, the forager can visit the same target site many times. In the case of destructive foraging, the target site found by the forager becomes undetectable in subsequent flights. Let $\lambda$ be the mean free path of the forager between two successive target sites (for two dimensions $\lambda=(2r_v\phi)^{-1}$ where $\phi$ is the target-site area density).

\begin{figure}[!htb]
  \includegraphics[width=14cm]{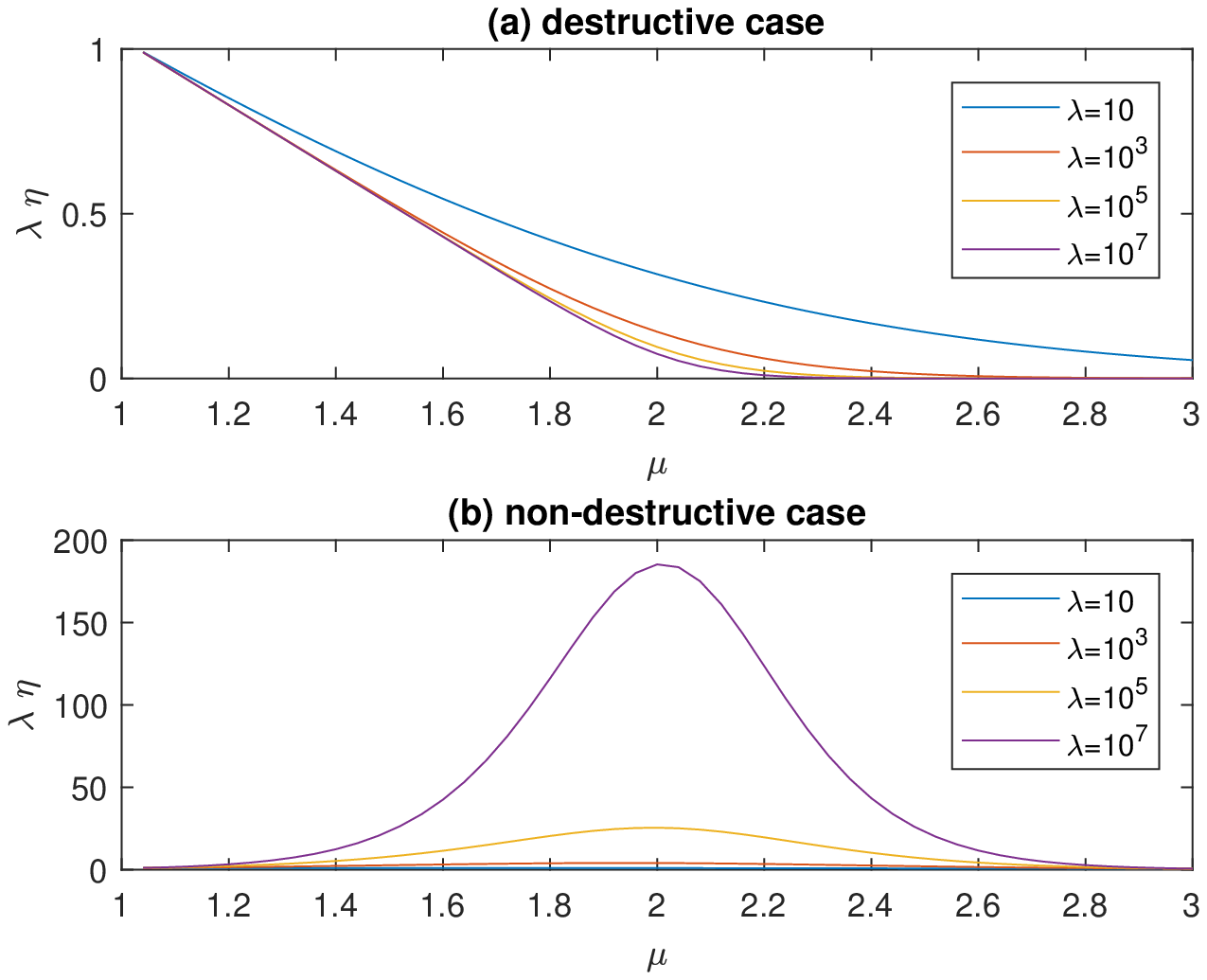}
  \caption{Searching efficiency of standard L{\'e}vy flight for different mean free path $\lambda$}
  \label{f12}
\end{figure}

On the basis of above behaviors, assume that the flight distance is distributed as the L{\'e}vy distribution
\begin{eqnarray}\label{lf}
p\left( l \right) \sim l^{-\mu},~l\ge r_v,~1<\mu<3.
\end{eqnarray}
As shown in Fig. \ref{f12}, researchers find that $\mu\approx 2$ and $\mu \to 1$ will result in an optimal searching efficiency for the non-destructive case and destructive case, respectively. For more details about the model and existing results, one may refer to the works of \cite{viswanathan1999optimizing,viswanathan2000Levy} and references therein.

\section{Searching efficiency with a tempered L{\'e}vy flight}\label{sec3}
\begin{figure}[!htb]
  \includegraphics[width=14cm]{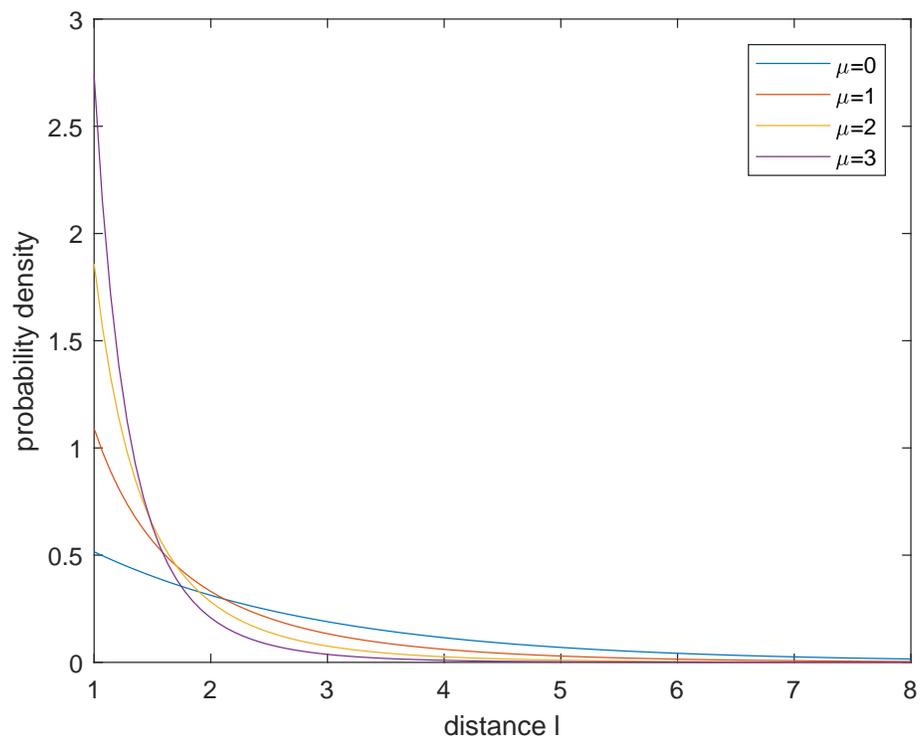}
  \caption{Probability density for tempered L{\'e}vy distribution with $\rho=0.5$ for different $\mu$}
  \label{f13}
\end{figure}

In almost all the existing literatures about L{\'e}vy flight foraging, it is assumed that the flight distance at each step is independently distributed as (\ref{lf}). Distribution (\ref{lf}) is power-law decaying, which indicates that a large jump length will appear more frequently compared with the traditional Gaussian distribution. In practical foraging, after the forager determines the flight distance at some step, the flight will be interrupted by some unknown reasons, such as obstacles on the flight direction, natural enemies in the vision distance, and restrictions in the energy storage for each flight. Because of these reasons, we can assume that the flight distance is distributed as
\begin{eqnarray}\label{tlf}
p\left( l \right) \sim {\rm e}^{-\rho l}{l^{ - \mu }},~{r_v} \le l,~\rho>0,~\mu \ge 1,
\end{eqnarray}
which indicates that the forager can keep the flight direction with the probability of an exponential distribution. Fig. \ref{f13} shows the probability density function (pdf) of a tempered  L{\'e}vy distribution and one can find that the density decreases slower with a smaller $\mu$, which means that a larger jump length is more likely to happen.

\textbf{Remark}
The difference between (\ref{lf}) and (\ref{tlf}) is that the power law distribution is tempered by an exponential decaying ${\rm e}^{-\rho l}$. The exponential part ${\rm e}^{-\rho l}$ can be viewed as the probability density that the forager can keep its flight direction before he completes one flight in the existence of some unknown factors and $\rho$ is determined by the environment. Because L{\'e}vy distribution is now tempered by ${\rm e}^{-\rho l}$, the first and second order moments of distribution (\ref{tlf}) exist for arbitrary $\mu\in \mathbb R$. In the paper, we will discuss the problem in a wider range $\mu\in [1,\infty)$ rather than $(1,3)$ for the L{\'e}vy distribution.

\subsection{The non-destructive case}
In this part, we will borrow the idea from \cite{viswanathan1999optimizing} to optimize the searching efficiency. Given the pdf of the flight distance as (\ref{tlf}), the mean flight distance can be calculated as
 \begin{eqnarray}\label{}
 \begin{array}{rl}
\left\langle l \right\rangle  =&\hspace{-6pt} \frac{{\int_{{r_v}}^\lambda  {{e^{ - \rho \left| x \right|}}{{\left| x \right|}^{ - \mu  + 1}}} dx + \lambda \int_\lambda ^\infty  {{e^{ - \rho \left| x \right|}}{{\left| x \right|}^{ - \mu }}} dx}}{{\int_{{r_v}}^\infty  {{e^{ - \rho \left| x \right|}}{{\left| x \right|}^{ - \mu }}} dx}}\\
 = &\hspace{-6pt} \frac{{ {{\Gamma _{up}}\left( {\rho {r_v},2 - \mu } \right) - {\Gamma _{up}}\left( {\rho \lambda ,2 - \mu } \right)} + \lambda \rho {\Gamma _{up}}\left( {\rho \lambda ,1 - \mu } \right)}}{{\rho {\Gamma _{up}}\left( {\rho {r_v},1 - \mu } \right)}}
\end{array}
\end{eqnarray}
where, the incomplete gamma function $\Gamma_{up}$ is defined as
 \begin{eqnarray}\label{mfl}
{\Gamma _{up}}\left( {x,a} \right) = \int_x^\infty  {{t^{a - 1}}{{\rm{e}}^{ - t}}{\rm{d}}t}.
\end{eqnarray}

Let $N$ be the mean number of flights taken by a L{\'e}vy forager while travelling between two successive target sites. Since the first and second order moments of tempered L{\'e}vy distribution exist, the trajectory of the forager will result in a Brownian motion. According to the existing results by \cite{viswanathan1999optimizing}, for the non-destructive case, it follows that the mean flight number between two successive targets can be estimated as
 \begin{eqnarray}\label{mfn2}
{N_n} \approx \left(\frac{{{\lambda ^2}}}{{2D}}\right)^{\frac{1}{2}}
\end{eqnarray}
where, $D$ is the diffusion constant. According to the standard diffusion equation in Section \ref{sec4}, it is found that the diffusion constant $D=\frac{a}{2b}$, where $a$ is the second order moment of flight distance and $b$ is the mean of the waiting time. Since we do not take the time into consideration, one can conclude that $N_n$ is proportional to $\left(\frac{\lambda^2}{a}\right)^{\frac{1}{2}}$. Here, $a$ can be calculated as
  \begin{eqnarray}\label{}
a = \int_{{r_v}}^\infty  {{{\rm{e}}^{ - \rho l}}{l^{2 - \mu }}{\rm{d}}l}  = {\rho ^{\mu-3 }}{\Gamma _{up}}\left( {\rho {r_v},3 - \mu } \right).
\end{eqnarray}
Based on the above analyses, we can then calculate the searching efficiency which is defined as
 \begin{eqnarray}\label{efficiency}
\eta  = \frac{1}{{{N}\left\langle l \right\rangle }}.
\end{eqnarray}

\begin{figure}[!htb]
  \includegraphics[width=14cm]{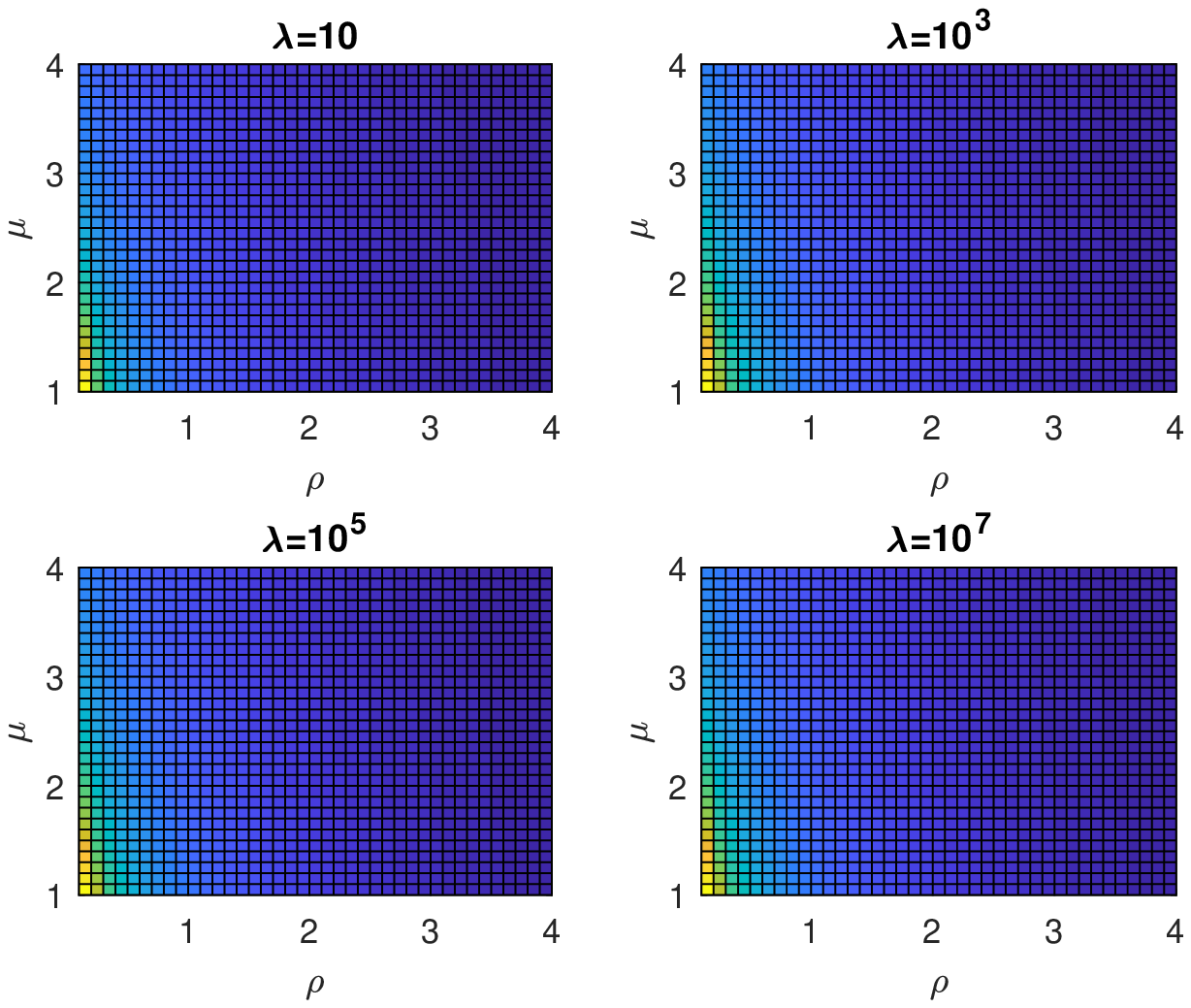}
  \caption{Searching efficiency $\eta \lambda$ for different order $\mu$ and $\rho$: the nondestructive case with different $\lambda$}
  \label{f1}
\end{figure}

Take $r_v$ as 1 when simulating and the results for different mean free path $\lambda$ are shown in Fig. \ref{f1}. Following observations can be drawn
\begin{itemize}
  \item [(1)] For fixed mean free path $\lambda$ and $\rho$, a smaller $\mu$ will result in a higher searching efficiency.
  \item [(2)] For fixed mean free path $\lambda$ and $\mu$, a larger $\rho$ will result in a higher searching efficiency.
  \item [(3)] The mean free path $\lambda$ almost has no influence on the choice of $\mu$ and $\rho$ to derive the highest searching efficiency.
\end{itemize}

As interpreted in the existing papers, the L{\'e}vy distribution can lead to a higher efficiency in a sparse area due to the higher probability of large jump lengths. For this issue, a smaller $\mu$ or $\rho$ will both decrease the decaying speed of the probability density, which means that the large jump lengths are more likely to appear. Hence, observations (1) and (2) can be explained since frequently large jump lengths can help covering a wider range where it is more likely to find a target in a sparse area. Generally, the density of target site is sparse in practice which means that $\lambda$ is usually large. Due to the exponential decaying of tempered L{\'e}vy distribution, the value of ${\lambda \int_\lambda ^\infty  {{e^{ - \rho \left| x \right|}}{{\left| x \right|}^{ - \mu }}} dx}$ is quite small and almost has no influence on the searching efficiency. It can then explain why the results of Fig. \ref{f1} with different $\lambda$ are similar.

One can also interpret the observations from the practical perspective. As discussed before, the tempered item ${\rm {e}}^{-\rho l}$ can be viewed as the probability density that the forager can keep its flight direction before he completes one flight in the existence of some unknown factors. Thus, a smaller $\lambda$ means that the probability of a forager to encounter some uncertain factors is lower and the foraging efficiency should be higher.


\subsection{The destructive case}
\begin{figure}[!htb]
  \includegraphics[width=14cm]{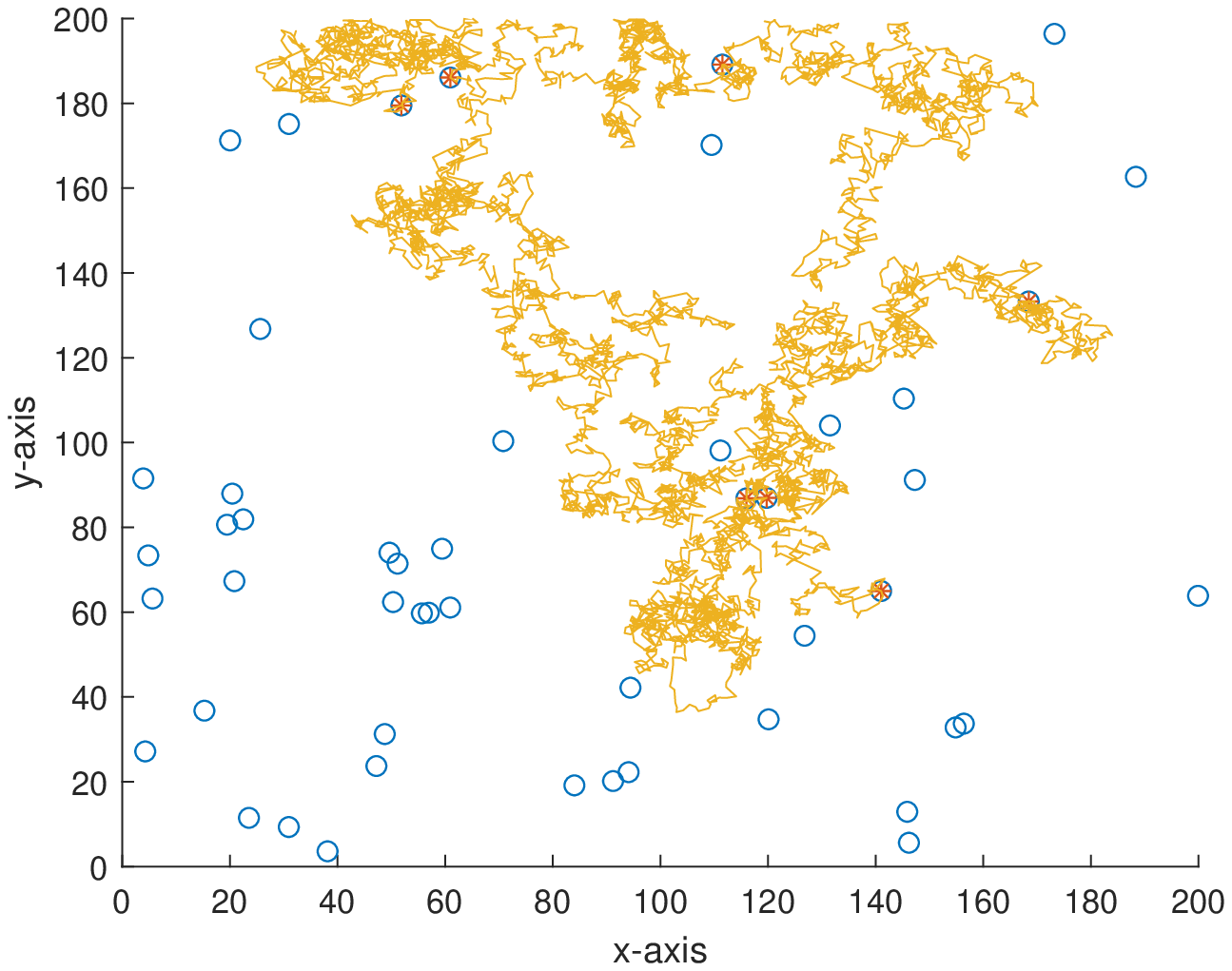}
  \caption{A typical example of foraging procedure with $\lambda=1$ and $\mu=1$}
  \label{f5}
  \end{figure}

For the destructive case, the mean number $N$ can be expressed as
 \begin{eqnarray}\label{dmfn2}
{N_d} \approx \frac{{{\lambda ^2}}}{{2D}}.
\end{eqnarray}
Similar to the non-destructive case, one can then calculate the searching efficiency using (\ref{efficiency}). The results are shown in Fig. \ref{f5}, which is very similar to the non-destructive case. It is found that a smaller $\mu$ or $\rho$ will both result in a higher search efficiency. The mean free path $\lambda$ almost has no influence on the optimal choice of parameters $\mu$ and $\rho$. We have shown that for the L{\'e}vy distribution, $\mu \rightarrow 1$, where a large jump length appears more likely, will lead to a higher searching efficiency. Thus, a smaller $\rho$ and $\mu$ will also result in a larger searching efficiency because large jump lengths are more likely to happen.

\subsection{Experimental results}
\begin{figure}[!htb]
  \includegraphics[width=14cm]{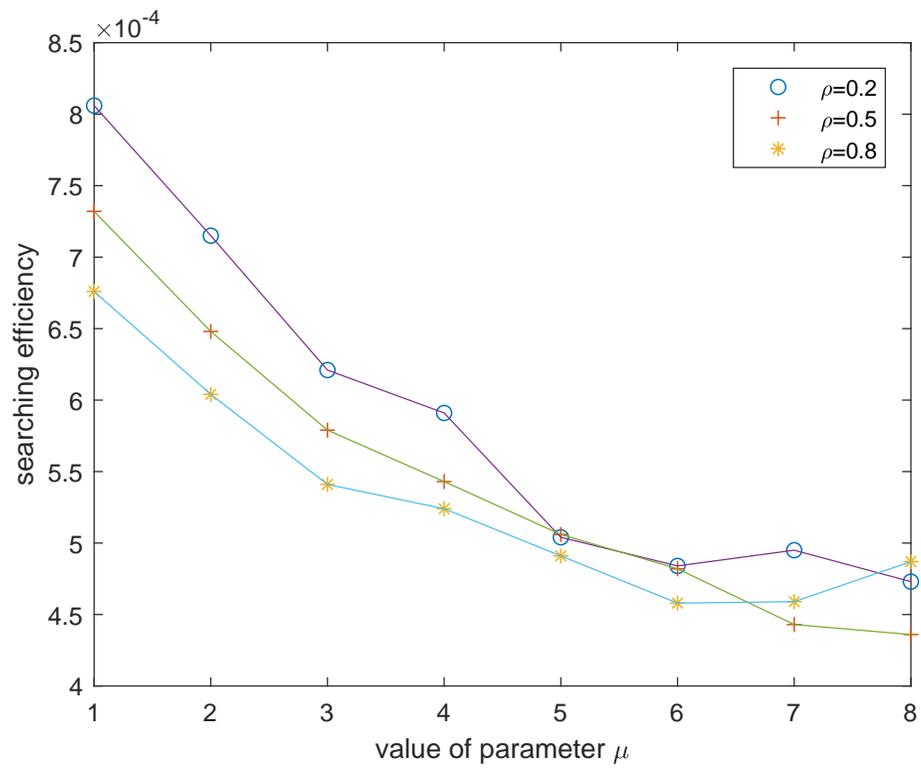}
  \caption{Experimental results of searching efficiency for different $\lambda$ and $\mu$}
  \label{f3}
  \end{figure}

\begin{figure}[!htb]
  \includegraphics[width=14cm]{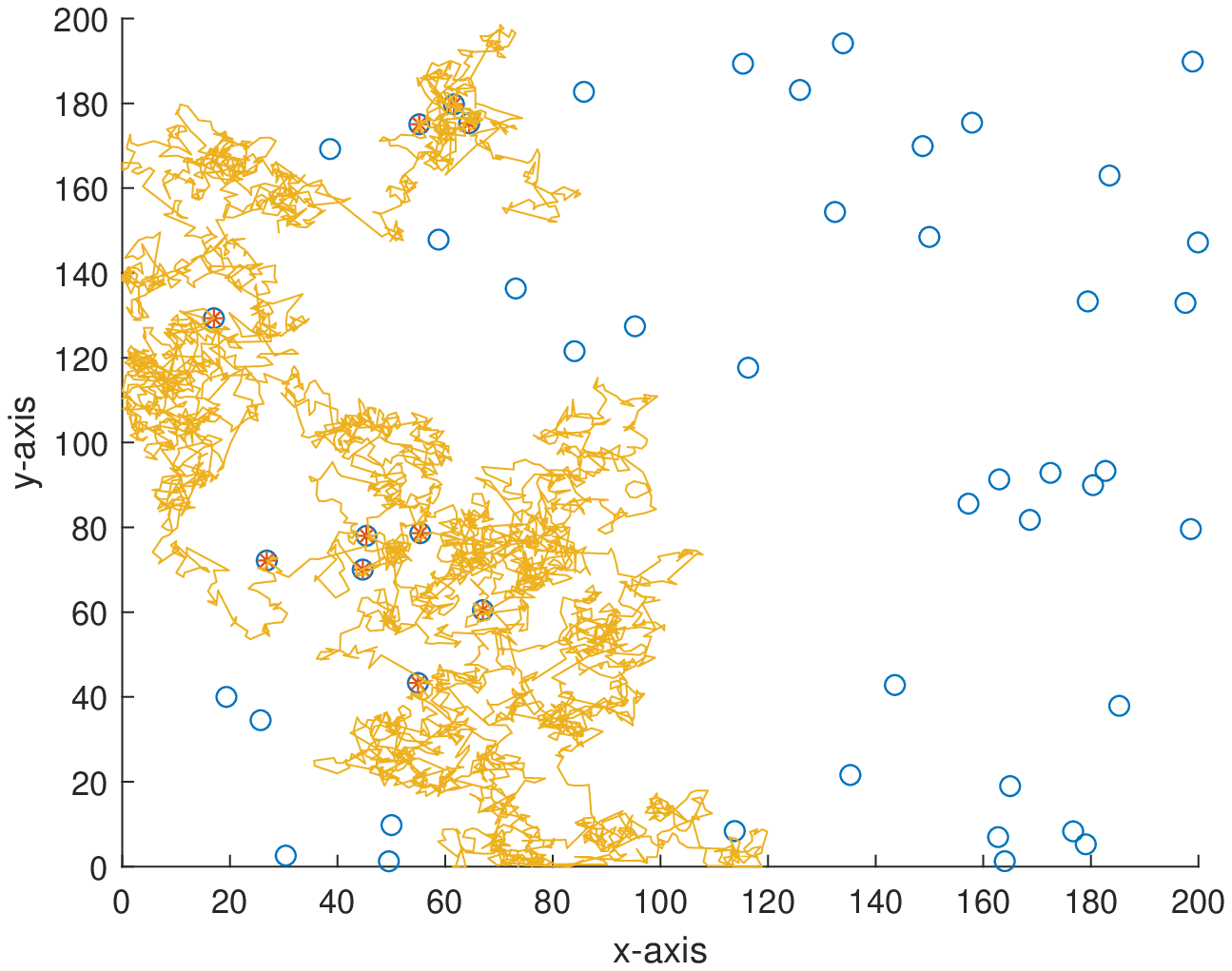}
  \caption{A typical example of foraging procedure with $\lambda=0.5$ and $\mu=1$}
  \label{f4}
  \end{figure}

\begin{figure}[!htb]
  \includegraphics[width=14cm]{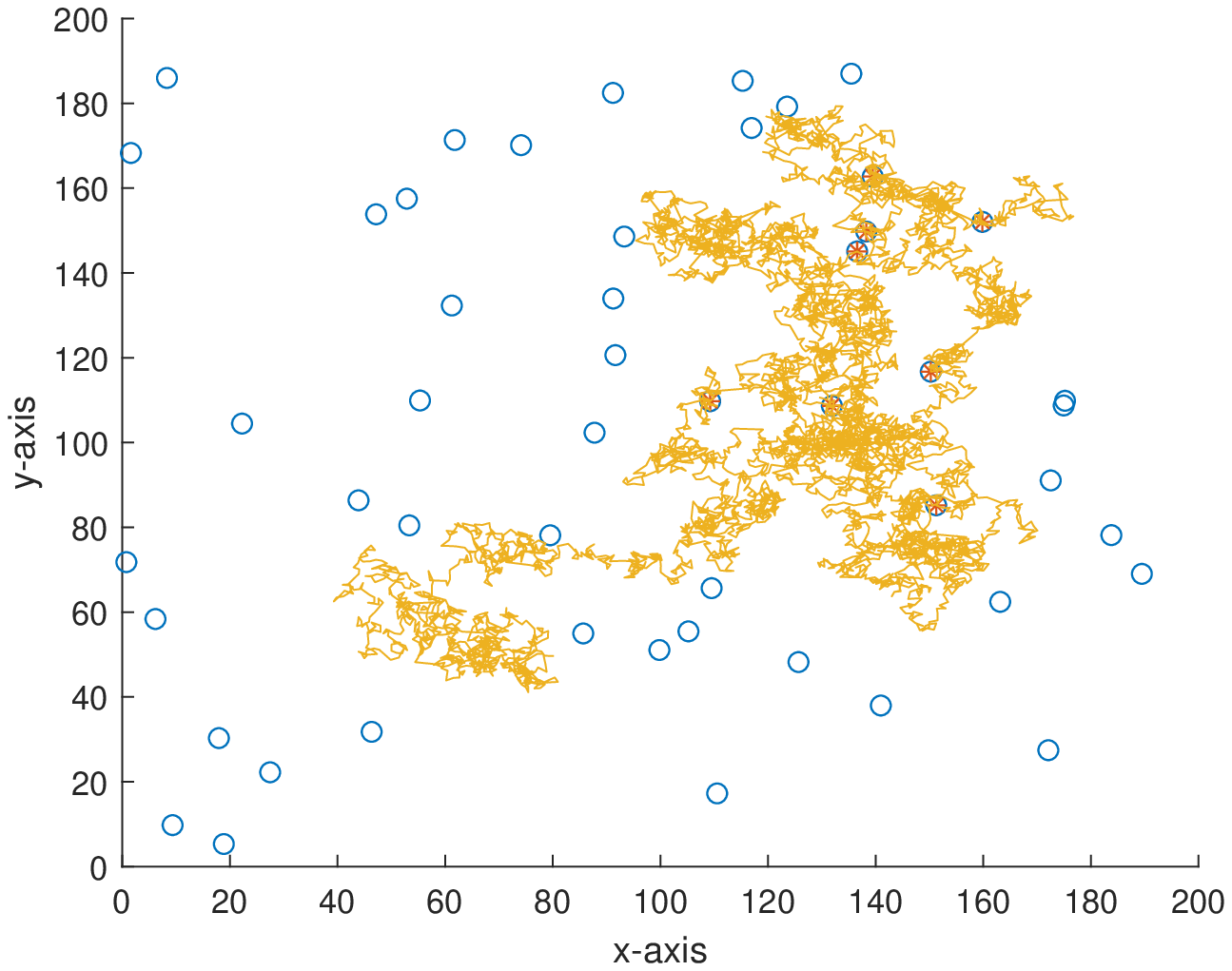}
  \caption{A typical example of foraging procedure with $\lambda=0.5$ and $\mu=3$}
  \label{f6}
  \end{figure}

We also implement an experiment for validate the theoretical analyses. Consider a $200\times 200$ area and $50$ targets are uniformly distributed in this area. The vision distance is $r_v=1$ and the total flight distance is no longer than $10000$ which can be viewed as the flight capability of the forager. The searching efficiency is estimated as $\frac{N_{num}}{L_{total}}$ where $N_{num}$ is the number of found targets and $L_{total}$ is the total flight distance. From Fig. \ref{f3} where the searching efficiency is derived by averaging $100$ independent runs, one can find that a smaller $\rho$ and $\mu$ will both lead to a higher searching efficiency, which is consistent with the theoretical analyses. Because a larger $\mu$ will make the density function decrease quickly, the range of jump lengths is then very tight. Thus, the searching efficiency is very close for a large $\mu$ where the jump lengths are all around the vision distance $r_v$. Fig. \ref{f4} - Fig. \ref{f6} give some typical foraging procedure for different parameters and one can find that all of them perform a Brownian motion. Additionally, larger jump lengths frequently appear in Fig. \ref{f4} compared with the other two figures, for which the searching efficiency is the highest.

\begin{figure}[!htb]
  \includegraphics[width=14cm]{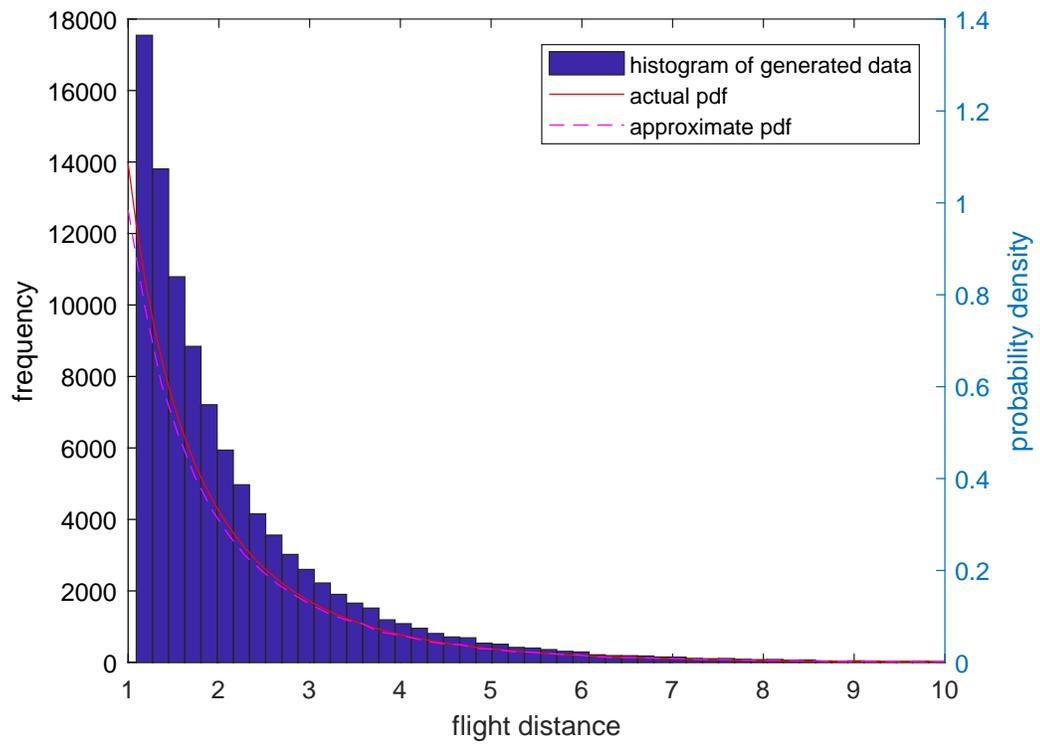}
  \caption{The actual tempered L{\'e}vy density function with $\rho=0.5$ and $\mu=1$ and the statistic results of generated jump lengths}
  \label{f7}
  \end{figure}

\textbf{Remark}: In this paper, we numerically generate the jump lengths distributed as a tempered L{\'e}vy distribution and Fig. \ref{f7} shows the actual density function and the statistic result of generated jump lengths. It is found that the statistic result is very close to the actual density function.

\section{Master equations}\label{sec4}
In the previous, we have not taken the flight time into consideration. Assume that the flight speed $v$ is constant during the foraging process and treat the flight time between two flights as the waiting time. Then, the pdf of waiting time is the same as the flight distance with a scaling parameter $v$, which can be expressed as
\begin{eqnarray}\label{}
p\left( t \right) \sim {\rm e}^{-\frac{\rho}{v} t}{t^{ - \mu }},~t\ge \frac{r_v}{v}.
\end{eqnarray}

Let us introduce the Fourier transform for the flight distance and the Laplace transform for the waiting time respectively as
\begin{eqnarray}\label{wk}
W\left( k \right) = \int_{{r_v}}^h {{{\rm{e}}^{ikl}}p\left( l \right){\rm{d}}l}
\end{eqnarray}
and
\begin{eqnarray}\label{ft}
\Psi \left( s \right) = \int_{{{{r_v}} \mathord{\left/
 {\vphantom {{{r_v}} v}} \right.
 \kern-\nulldelimiterspace} v}}^{{h \mathord{\left/
 {\vphantom {h v}} \right.
 \kern-\nulldelimiterspace} v}} {{{\rm{e}}^{ - st}}p\left( t \right){\rm{d}}t} .
\end{eqnarray}

The famous Montroll-Weiss equation \citep{montroll1965random} in Fourier-Laplace space is in the following form
\begin{eqnarray}\label{mw}
P\left( {k,s} \right) = \frac{{1 - \Psi \left( s \right)}}{s}\frac{1}{{1 - W\left( k \right)\Psi \left( s \right)}}.
\end{eqnarray}
 Now consider the extreme distribution of $W(k)$ and $\Psi(s)$ with $k\to 0$ and $s\to 0$, respectively. It is followed that
\begin{eqnarray}\label{aft}
\begin{array}{rl}
\Psi \left( s \right) = &\hspace{-6pt}\int_{{{{r_v}} \mathord{\left/
 {\vphantom {{{r_v}} v}} \right.
 \kern-\nulldelimiterspace} v}}^{{\infty}} {{{\rm{e}}^{ - st}}p\left( t \right){\rm{d}}t} \\
 = &\hspace{-6pt}\int_{{{{r_v}} \mathord{\left/
 {\vphantom {{{r_v}} v}} \right.
 \kern-\nulldelimiterspace} v}}^{{\infty}} {\left( {1 - st + o\left( {{s}} \right)} \right)p\left( t \right){\rm{d}}t} \\
 =&\hspace{-6pt} 1 - bs + o\left( {{s}} \right),
\end{array}
\end{eqnarray}
where, $b$ is the mean of flight time $t$ and $o(\cdot)$ means the higher order infinitesimal. In the following, we will present two different types of master equations for this foraging procedure.
\subsection{The standard diffusion equation case}\label{sub1}
Assume that the searching direction $\theta$ is uniformly distributed in the interval $[0,2\pi)$. If the waiting time and the flight distance are independent, then the location of the forager can be formulated as $(x,y)=(l\cos\theta,l\sin\theta)$ and the following equation holds
 \begin{eqnarray}\label{awk}
\begin{array}{l}
W\left( k \right) = \frac{1}{{2\pi }}\int_0^{2\pi } {\int_{{r_v}}^{\infty} {{{\rm{e}}^{il\left( {{k_1}\cos \theta  + {k_2}\sin \theta } \right)}}p\left( l \right)} } {\rm{d}}l{\rm{d}}\theta \\
 = \frac{1}{{2\pi }}\int_0^{2\pi } {\int_{{r_v}}^{\infty} {\left( {1 + il\Theta  + {{\left( {il\Theta } \right)}^2} + o\left( {{\Theta ^2}} \right)} \right)p\left( l \right)} } {\rm{d}}l{\rm{d}}\theta \\
 = 1 + \frac{1}{{2\pi }}\int_0^{2\pi } {\int_{{r_v}}^{\infty} {{{\left( {il\Theta } \right)}^2}p\left( l \right)} } {\rm{d}}l{\rm{d}}\theta  + o\left( {{\Theta ^2}} \right)\\
 = 1 + \frac{a}{2}\left( {{{\left( {i{k_1}} \right)}^2} + {{\left( {i{k_2}} \right)}^2}} \right) + o\left( {{\Theta ^2}} \right)
\end{array}
\end{eqnarray}
where, $\Theta=k_1\cos \theta+k_2\sin \theta$ and $a$ is the second order moment of the flight distance.

Substitute(\ref{aft}) and (\ref{awk}) into the Montroll-Weiss equation and ignore the higher order infinitesimal, yielding,
\begin{eqnarray}\label{ms}
\begin{array}{rl}
P\left( {k,s} \right) =&\hspace{-8pt} \frac{b}{{bs - \frac{a}{2}(ik_1)^2-\frac{a}{2}(ik_2)^2}} \\
= &\hspace{-8pt} \frac{1}{{s -\frac{a}{2b}\left((ik_1)^2+(ik_2)^2\right)}}.
\end{array}
\end{eqnarray}
Perform inverse Fourier-Laplace transform and one can derive the master equation
\begin{eqnarray}\label{mst}
\frac{\partial }{{\partial t}}p\left( {x,y,t} \right) = \frac{a}{2b}\frac{{{\partial ^2}}}{{\partial {x^2}}}p\left( {x,y,t} \right)+\frac{a}{2b}\frac{{{\partial ^2}}}{{\partial {y^2}}}p\left( {x,y,t} \right).
\end{eqnarray}

\textbf{Remark}
Unlike the master equation derived by \cite{zeng2014optimal}, the master equation in this study is a normal diffusion equation since the first and second order moments exist. We have to mention that the master equation proposed by \cite{zeng2014optimal} should also be standard diffusion equation rather than fractional diffusion differential equation since the L{\'e}vy distribution is truncated by the mean free path $\lambda$. Moreover, the master equation should be two-dimensional rather than one-dimensional.

\subsection{The tempered fractional diffusion equation case}
In this subsection, our purpose is to express the master equation as a tempered fractional diffusion equation and we have restrict $\mu$ varies from 1 to 2 to derive the tempered fractional derivative expression. The vector jump length can be described as $l\Theta$, where $\Theta=(\cos\theta,\sin\theta)$. From equation (7.9) in the book of \cite{meerschaert2012stochastic}, it shows that
\begin{eqnarray}\label{tjl}
\begin{array}{rl}
W\left( k \right) =&\hspace{-6pt} \int_{\left\| \Theta  \right\| = 1} {\int_{{r_v}}^\infty  {{{\rm{e}}^{ik \cdot l\Theta }}p\left( l \right){\rm{d}}lM\left( {{\rm{d}}\Theta } \right)} } \\
 =&\hspace{-6pt} 1 + \int_{\left\| \Theta  \right\| = 1} {\int_{{r_v}}^\infty  {\left( {{{\rm{e}}^{ik \cdot l\Theta }} - 1} \right)p\left( l \right){\rm{d}}lM\left( {{\rm{d}}\Theta } \right)} } \\
 =&\hspace{-6pt} 1 + C\int_{\left\| \Theta  \right\| = 1} {\left[ {{{\left( {\lambda  - ik \cdot \Theta } \right)}^{\mu  - 1}} - {\lambda ^{\mu  - 1}}} \right]} M\left( {{\rm{d}}\Theta } \right),
\end{array}
\end{eqnarray}
where $k\cdot \Theta=k_1\cos\theta+k_2\sin \theta$, $M({\rm d}\Theta)$ is a uniform distribution on a unit circle, and $C$ is a constant relevant to coefficients $\rho$ and $\mu$.

Substitute(\ref{aft}) and (\ref{tjl}) into the Montroll-Weiss equation (\ref{mw}) and ignore the higher order infinitesimal, yielding,
\begin{eqnarray}\label{me2}
P\left( {k,s} \right) = \frac{1}{{s - \frac{C}{b}\int_{\left\| \Theta  \right\| = 1} {{{\left( {\lambda  - ik \cdot \Theta } \right)}^{\mu  - 1}} - {\lambda ^{\mu  - 1}}} M\left( {{\rm{d}}\Theta } \right)}}.
\end{eqnarray}

Define
\begin{eqnarray}\label{}
{}^\lambda \nabla _M^\alpha f\left( x \right) = \int_{\left\| \Theta  \right\| = 1} {{}^\lambda D_M^\alpha } f\left( x \right)M\left( {{\rm{d}}\Theta } \right)
\end{eqnarray}
where,
\begin{eqnarray}\label{}
{}^\lambda D_\Theta ^\alpha f\left( x \right)= \frac{\alpha }{{\Gamma \left( {1 - \alpha } \right)}}\int_0^\infty  {\left[ {g\left( t \right) - g\left( {t - r} \right)} \right]{e^{ - \lambda r}}{r^{ - \alpha  - 1}}{\rm{d}}r}
\end{eqnarray}
with $g\left( t \right) = f\left( {x + t\Theta } \right)$ is the generator form for vector tempered fractional derivative.

Inverse (\ref{me2}) to derive the master equation
\begin{eqnarray}\label{me333}
\frac{{\partial p\left( {L,t} \right)}}{{\partial t}} = \frac{C}{b}{}^\lambda \nabla _M ^{\mu  - 1}p\left( {L,t} \right),
\end{eqnarray}
where, $L$ is a vector $(x,y)$.

\textbf{Remark}
Since the first order and second order moments of tempered L{\'e}vy distribution exist, the resulting standard diffusion equation (\ref{mst}) makes sense. Interestingly, we borrow the idea from \cite{meerschaert2012stochastic} and give another expression of the master equation, where vector tempered fractional derivative is used. In this paper, we do not give detailed proof for the derivation of vector tempered fractional derivative and one can refer to Chapter 6 and 7 in the book of \cite{meerschaert2012stochastic}. All these indicate that tempered fractional diffusion equation is in fact a different expression of the standard diffusion.

\section{Conclusion}\label{sec5}
In this paper, we consider the optimal random foraging whose flight distance is distributed according to a tempered L{\'e}vy distribution $p(l)\sim {\rm e}^{-\rho l}l^{-\mu}$. It is found that a higher searching efficiency can be derived when we choose a smaller $\rho$ or $\mu$, which results in a slower decaying speed. Furthermore, we obtain the master equation of the random foraging. A standard diffusion equation is derived since the first and second order moments of the distribution for flight distance exist. Using the definition of tempered fractional derivative, a vector tempered fractional diffusion equation is then derived, which can be viewed as a special expression for the standard diffusion. A promising research topic can be directed to finding the optimal searching strategy for other types of flight distance distributions.

\section*{Conflict of Interest Statement}
The authors declare that there is no conflict of interests regarding the publication of this paper

\section*{Author Contributions}
Yuquan Chen mainly contributed to the theoretical analysis and accomplishing the paper. Derek Hollenbeck mainly contributed to the numerical simulation. Yong Wang and YangQuan Chen contributed for providing the idea of using tempered L{\'e}vy distribution in foraging and helped revising the paper.

\section*{Funding}
 This work was fully supported by China Scholarship Council (No. 201706340089) and NSF NRT Fellowship.

%


\bibliographystyle{model1-num-names} 
\bibliography{ICFDAdatabases}


\section*{Figure captions}

\begin{figure}[!htb]
  \includegraphics[width=14cm]{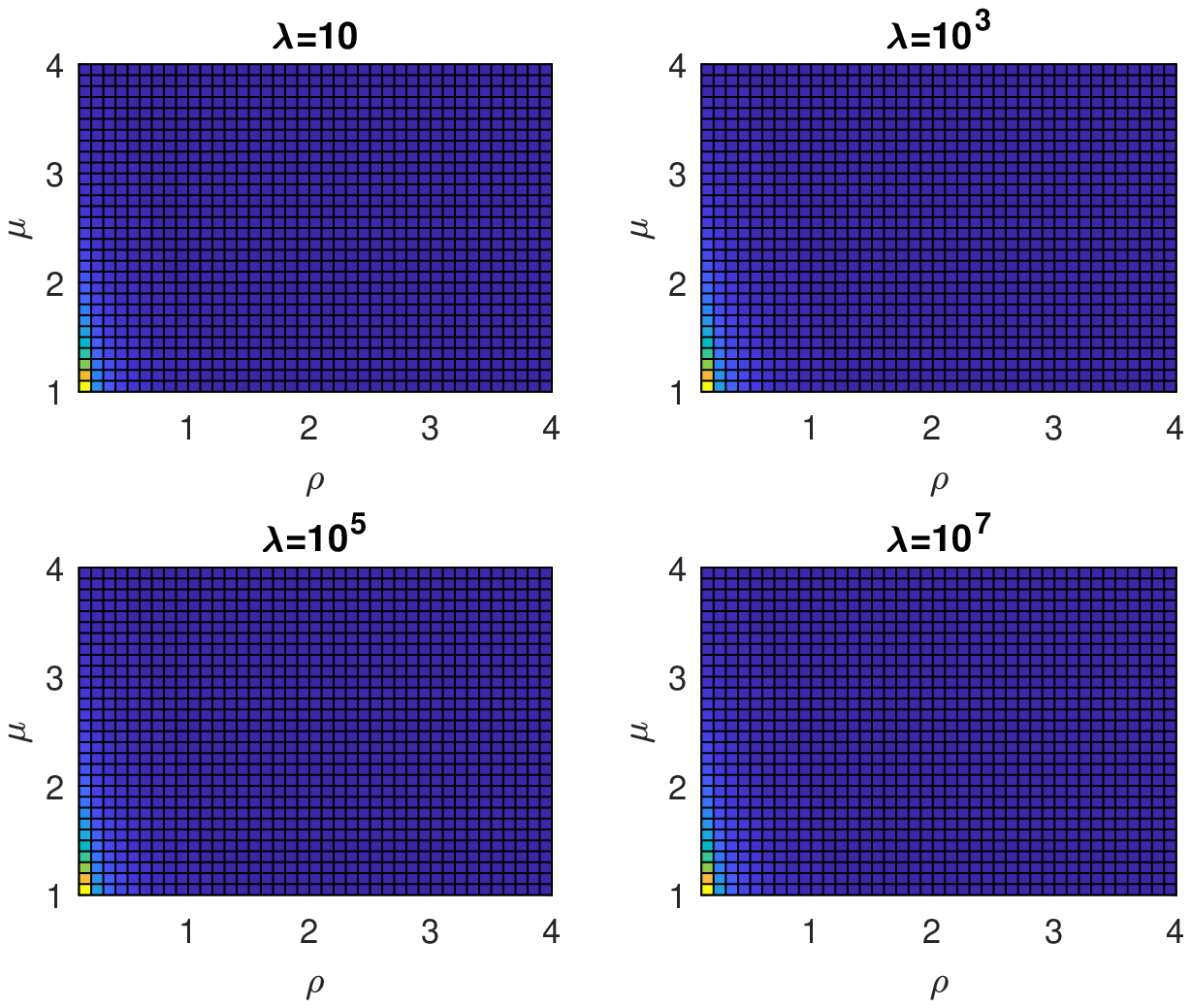}
  \caption{Searching efficiency $\eta \lambda$ for different order $\mu$ and $\rho$: the destructive case with different $\lambda$}
  \label{f2}
  \end{figure}

\end{document}